\documentstyle[12pt]{article}

\def\be{\begin{equation}}
\def\ee{\end{equation}}

\begin{document}

\begin{center}
{\Large \bf SELF-FORCES IN THE SPACETIME}
\end{center}
\begin{center}
{\Large\bf OF MULTIPLE COSMIC STRINGS}
\end{center}

\begin{center}
E.R. Bezerra de Mello\footnote{E-mail address: 
eugenio@dfjp.ufpb.br}, 
V.B. Bezerra\footnote{E-mail address: valdir@fisica.ufpb.br}, 
\end{center}
\begin{center}
and Yu.V. Grats\footnote{On
leave from M.V. Lomonosov Moscow State University, Moscow, Russia}
${^,\ }$\footnote{E-mail address: grats@grg1.phys.msu.su}
\end{center}

\begin{center}
Departamento de F\'{\i}sica, Universidade Federal da Para\'{\i}ba,\\
Caixa Postal 5008, CEP 58051-970 Jo\~{a}o Pessoa, Pb, Brazil
\end{center}

\begin{abstract}
We calculate the electromagnetic self-force on
a stationary linear distribution of four-current in the spacetime 
of multiple cosmic strings.
It is shown that if the current is infinitely thin and 
stretched along a line which is
parallel to the
strings the problem admits an explicit solution.\\ 

\end{abstract}
\newpage

\section{\bf Introduction}

The investigations in quantum and classical field 
theory in curved spacetime have shown that 
in many cases the local observables of the theory
can not be expressed in terms of local geometry, but depend on the
global structure of the space.
This fact was interpreted as the
manifestation of nonlocal (topological) influence of gravity 
on matter fields. 
The further researches have revealed that some of these nonlocal
processes are not of pure theoretical interest,
but may play an important role in modern particle physics, 
astrophysics, and cosmology [1, 2].
>From the general point of view, nonlocal aspects are of fundamental
importance in describing the physics of a given system. 

Spaces with conical singularities are the new object of
such an investigation. As a rule these spaces are associated with
cosmic string spacetimes. 
The detailed survey of the problems
associated with cosmic strings and an extensive list of references
can be found in [3, 4].
As a matter of fact spaces with similar features appear in some other
important physical applications. It is well 
known, for example, that some of the
linear defects in crystalls, namely disclinations, are  
described geometrically as linear conical defects in a locally flat  
space [5]. Conical singularities appear in the off-shell
calculations near the horizon in the Euclidean section of 
the Schwarzchild spacetime [6], and in connection with thermodynamics
in the presence of cosmological horizons [7]. Some of conical spaces
possess a trivial local geometry and this
enables one to expose pure nonlocal effects in the
gravitational interaction.
In the particular case of the static straight infinitely thin
cosmic string the spacetime looks like the direct product 
$M_2\times Cone$\ of the two-dimensional Minkowski space and a cone. 
The corresponding Riemann tensor vanishes everywhere except on the
symmetry axis, where it has a $\delta$ - like
singularity [8].
So, infinitely thin cosmic strings
do not affect the local geometry of the spacetime but change its
global properties, and therefore their effect on the matter fields 
is purely topological. 

In recent years, the interest in cosmic strings has been associated
with investigations in the quantum field theory mainly, 
and a lot of interesting
quantum effects have been studied. The vacuum expectation
values of the energy-momentum tensor for fields of different
spins have been calculated and the phenomenon of vacuum polarization
was considered in detail [9]. These calculations have also been extended to the
case of finite temperatures [10].
Recently it was shown that the presence of other
strings modifies the spectrum of vacuum fluctuations of a scalar field 
in such a way that a Casimir-like interaction
arises between the strings, and two parallel cosmic strings are
attracted to each other with a force that decreases with third power of 
the distance
between the strings [11]. Later the results
of the paper [11] were used to study the vacuum polarization
of an electromagnetic field [12].

As it was mentioned above, the effect of the straight string on quantized
fields is purely topological. One often says that it is due to the
deformation of a Green function by the nontrivial global structure
of the space. It is
clear that similar effects must take place in the classical
theory too if we consider both point particles and
classical fields. Indeed in 
quantum and classical field theories we work with Green functions
which are the solutions
of the same equation. So, if some distortion of one of the functions,
say Feynman one, is found, we must expect that the others will 
be distorted too. 
In fact
some nontrivial classical effects which are due to the global
distinction of the conical space from the Minkowski one have been  
found. The repulsion
of a charged test particle from a cosmic string is an example of this
type of effects [13]. A similar effect was discovered for a gravitating
pointlike particle [14]. In the case of the linear distribution
of a current the situation is somewhat more interesting because
both electric and magnetic self-forces are induced [15].

In all the works mentioned previously, the effects were studied in the
spacetime of a single linear defect. In typical cosmological
scenarios, however, strings are generated in the form of 
random network of straight moving segments. Computer simulations
show [16] that there are about 
ten straight cosmic strings under the Hubble
horizon at our era. Thus, the multistring spacetime is much more
appropriate model for the real spacetime than the particular case of only
one cosmic string.
Another motivation comes from the solid state physics because
the number of disclinations per unit volume 
in a real crystal exceeds the unity by far, and
for this reason, it is necessary to consider spaces with two and more
conical singularities. This problem is much 
more complicated than the problem with only one linear defect.
Indeed the spacetime of a straight infinitely thin string possesses
four Killing vectors, and three of them,
namely $\partial/\partial t\ ,\
\partial/\partial z$\ and\  
$\partial/\partial\phi$\, give us the possibility to
separate the variables in the field equation.
The loss of the azimuthal symmetry, as it is in the multistring
case, makes impossible to proceed along this line.
One possible approach to this problem 
is associated with the use 
of perturbation theory [11], another one is based on the
possibility to solve some of the problems explicitly  
in low-dimensional gravity [17].

Here we present an explicitly solvable problem in the
(3+1)-dimensional multistring spacetime. To our knowledge this is
the first explicitly solvable nonlocal problem
which has been considered on the multistring space till now.
Our paper completes the investigations of topological effects
in multistring spaces that were initiated in the papers [11, 18, 19]. 

The paper is organized as follows. In Sec.2 we develop formal
expressions for the self-energy and self-force on a four-current
placed in a spacetime which is the direct product of the two-dimensional
Minkowski space and a Riemannian surface.  
The regularization procedure is described in Sec.3. 
In Sec.4 we apply the results obtained in previous sections in
order to consider the effect of topological self-action on a
current in the spacetime of multiple cosmic strings.
In Sec.5 we add some discussion and conclusion remarks about our
results and their possible application.

\section{\bf Formal expressions for self-energy and self-force}

Let us start with the metric of background spacetime which is
supposed to be the direct product of the two-dimensional Minkowski
space and a Riemannian surface. This is given by the interval
\be
ds^2 = dt^2 - dz^2 - \gamma_{ab} (x^c) dx^{a}dx^{b}\ ,\qquad
 a, b, c = 1, 2 \ .
\ee
The multistring space is a particular
case of the solution (1) when the section
$t=const, z=const$\ 
is a locally flat surface with a set of conical singularities [20].

On this background we shall consider a static electromagnetic
field which is
associated with a stationary distribution of a four-current of the form
\be
j^{\mu}(x_1) = (J^t,0,0,J^z)\frac{\delta^2 (x_1-x)}{\sqrt{\gamma(x)}}\ ,
\ee
where the components $J^t$\ and $J^z$\ are the linear charge density and
the electric current which are measured by an inertial observer with a
fixed $z$\ coordinate. 
 
The spacetime under consideration possesses a global timelike Killing
vector $\zeta_{(t)}=\partial/\partial t$\ . This enables us to
determine a total energy of the field. 
We consider the infinite current parallel to the strings, and
the spacetime which metric is invariant under
translations along the $z$ axis. So, the energy per unit length along
the current is independent on $z$\ , and the total energy is infinite.
But the linear density is well defined, and corresponding expression 
reads
\be
\frac{E}{\int dz}=\int d^2 x\sqrt{\gamma(x)} T_t^t(x)\ .
\ee

In Eq.(3) $T^t_t$ is the $tt$-component of the electromagnetic energy-momentum
tensor. To proceed further let us consider the $T^t_t$ - component taking
into account the symmetries of the spacetime and the current distribution.

In the Lorentz gauge $\nabla_{\nu}A^{\nu}=0$\ the components of a 
four-potential satisfy the
equation
\be
\nabla_{\nu}\nabla^{\nu}A_{\mu}-R_{\mu}^{\nu}A_{\nu}=
4\pi j_{\mu}\ ,
\ee
where $R_{\mu}^{\nu}$\ is the Ricci tensor. In the case of multistring
space it has a $\delta$-like singularities at the tops of the cones,
and it seems that we have to solve the wave equation with a singular
potential.

Fortunately 
the metric and the distribution of current are both invariant under
translations along $t$ and
$z$. This invariance and the residual freedom in the gauge transformations
enables us to choose a gauge in which the components of the four-potential
depend on the coordinates on the Riemannian surface only and 
$A_1=0=A_2$\ .

In this gauge the $tt$-component of the energy-momentum tensor
has the following form
\be
T_{tt}=\frac{1}{8\pi}\gamma^{ab}(\partial_a A_t\partial_b A_t
+ \partial_a A_z\partial_b A_z)\ .
\ee

Then, $R_a^b\ (a, b=1, 2)$\
are the only components of the Ricci tensor which are not equal to
zero in our coordinate system.
So, in the gauge above the second term in the lhs of the
Eq.(4) is equal to zero, and we get
\be
A_{\mu}(x_1)=-4\pi \int d^2 x_2\sqrt{\gamma(x_2)} 
G^{(2)}(x_1, x_2)j_{\mu}(x_2)\ .
\ee
In the last expression $G^{(2)}(x_1, x_2)$\ stands for the Green function
fulfilling the two-dimensional Poisson equation  
\be
\frac{1}{\sqrt{\gamma(x_1)}} \partial_{1a}\Biggl(\sqrt{\gamma(x_1)}
\gamma^{ab}(x_1)
\partial_{1b}
G^{(2)}(x_1, x_2)\Biggr)=\delta^2 (x_1, x_2)\ ,
\ee
where $\delta^2 (x_1, x_2)$\ is a covariant two-dimensional $\delta$ - function
and $\partial_{1b}$\ stands for the partial derivative 
$\partial/\partial x_1^b$\ .

So, for the distribution of current (2)
all the nonzero components of the four-potential can be expressed in terms
of the two-dimensional Green function (7). We shall see later that
this enables one to solve the problem of topological self-action
explicitly in much more 
general cases that it has been performed before.

Substituting (6) into (5) we get
\be
\int d^2 x\sqrt{\gamma(x)} T_t^t(x) = -2\pi\int d^2 x_1 d^2 x_2\sqrt{\gamma(x_1)} 
\sqrt{\gamma(x_2})\times
\ee
$$\Biggl(j_t(x_1)G^{(2)} (x_1, x_2)j_t(x_2)+
j_z(x_1)G^{(2)} (x_1, x_2)j_z(x_2)\Biggr)\ .
$$

The expression above is valid for any distribution of four-current, which is
stretched along $z$-axis and depends on the coordinates on the Riemannian
surface $x^a$\ only.
For the infinitely thin distribution of current 
(2) we obtain from it
\be
\frac{E_{stat}}{\int dz}=-2\pi \Biggl( J_t^2+J_z^2\Biggr)G^{(2)}(x, x)\ .
\ee

Equation (9) is the energy of a static electromagnetic
field per unit length along the strings. But an observer measures
the force, and corresponding expression is needed. In the case
of the pure electrostatic self-forces the situation is more or less
clear [15], the force which is measured in a locally inertial frame
can be obtained by taking the negative gradient of the self-energy.
In the case of magnetic force, as it is not a potential 
one, another procedure must
be adopted. So let us start from the force itself.

In the case
of infinitely thin current and for our particular geometry, 
the force per unit
length of the current measured by a locally inertial observer is
well defined. In an arbitrary coordinate system it is given by the relation
\be
f_{\mu}=\int d^2 x\sqrt{\gamma(x)} F_{\mu\nu}(x)j^{\nu}(x)\ .
\ee

In terms of the two-dimensional Green function this expression 
has the following form
\be
f_{\mu}=-4\pi\int d^2 x_1 d^2 x_2\sqrt{\gamma(x_1)}\sqrt{\gamma(x_2)}
\Biggl(j_{\nu}(x_1)j^{\nu}(x_2)\partial_{1\mu} G^{(2)}(x_1, x_2)- 
\ee
$$j_{\mu}(x_2)j^{\nu}(x_1)\partial_{1\nu}G^{(2)}(x_1, x_2)\Biggr)\ .$$
Integrating by parts the second term in the integrand and using the
current conservation law, we obtain
\be
f_{\mu}=-4\pi\int d^2 x_1 d^2 x_2\sqrt{\gamma(x_1)}\sqrt{\gamma(x_2)}
j_{\nu}(x_1)j^{\nu}(x_2)\partial_{1\mu}G^{(2)}(x_1, x_2)\ .
\ee  
Taking into account the explicit form of the current distribution 
(2) we get the following formal result
\be
f_{\mu}=-4\pi J^2\partial_{\mu}G^{(2)}(x, x) \ ,
\ee
where $J^2=(J_t^2-J_z^2)$\ is the squared invariant amplitude of the
current.
  
We are in a position to calculate the self-energy and self-force now,
but it is necessary to solve two problems.
Indeed both expressions for the self-energy (9) and for the self-force
(13) diverge because of the divergence of the Green function and its
derivatives in the coincidence limit. So, it is necessary to find the solution 
of the Poisson equation (7).
After that we have to adopt some regularization procedure and 
calculate the regularized values
of the Green function 
and its derivatives in the coincidence limit.

\section{\bf Green function of the Poisson equation\newline and its 
derivatives}

As it was shown by one of the authors [17]
the problem of finding $ G^{(2)} (x_1,x_2) $ is greatly simplified by the
fact that any two-dimensional Riemannian surface is locally conformal to
the Euclidean plane. This means that the coordinate system can be introduced,
in which in the vicinity of any point the metric takes the form
\be
\gamma_{ab}(x^c) = e ^{-\Omega (x^c) }\delta_{ab}\ ,\qquad  a, b, c=1, 2\ .
\ee

The primary goal of such a choice is to simplify the equation for
the Green function as far as possible.
Indeed in conformal coordinates the two-dimensional Poisson equation
(7)  
reduces to the one which has exactly the same form as the
Poisson equation on the Euclidean plane 
\be
\Delta_E G^{(2)}(\vec x_1, \vec x_2)=\delta^2 (\vec x_1-\vec x_2)\ ,
\ee
where $\Delta_E$ is the Laplace operator on the Euclidean plane.

The solution of (15) is given by
\be
G^{(2)}(\vec x_1, \vec x_2)=  \frac{1}{4\pi } \log |\vec x_1 - \vec x_2 |^2 +
\ee 
\begin{center}
an arbitrary analytic function of $(\vec x_1-\vec x_2)$\ ,
\end{center}
where $|\vec x_1-\vec x_2|$ denotes the Euclidean norm of the conformal vector
$(\vec x_1-\vec x_2)$\ .

The choice of the analytic function in (16) depends on the boundary conditions.
It seems to be impossible to formulate these conditions in the 
general case of Riemannian surfaces. So, let us consider the 
interesting case of
an infinite two-dimensional surface which is covered by the 
conformal coordinates globally. It is reasonable to demand that the
field of a point source tends to zero at the infinity. 
This boundary
condition reduces the choice of the analytic function 
to an orbitrary
constant which can be taken equal to zero without loss of
generality. Thus the Green function amounts to
\be
G^{(2)}(\vec x_1, \vec x_2)=  \frac{1}{4\pi } \log |\vec x_1 - \vec x_2 |^2\ .
\ee

Under the above assumptions this is
an explicit solution.
It is necessary to emphasize that 
in contrast with all the previous papers no symmetries of the 
two-dimensional surface were used to obtain it.
 
We are interested in the behaviour of the Green function and its
derivatives in the coincidence limit. In this limit both these
quantities diverge. To obtain a regularized value of the Green function
it is necessary to subtract from (17) the Green function on the 
Euclidean plane [1, 2]
\be
G^{(2)}_{reg}(\vec x_1, \vec x_2) = G^{(2)}(\vec x_1, \vec x_2)-
G_E^{(2)}(\vec x_1, \vec x_2)\ ,
\ee
where $G^{(2)}_E(\vec x_1, \vec x_2)$\ is the Euclidean Green function, which
can be written as
\be
G_E^{(2)}(\vec x_1, \vec x_2)=\frac{1}{4\pi}\log 2\sigma(\vec x_1, \vec x_2)\ .
\ee
In the expression above $\sigma(\vec x_1, \vec x_2)$\ is the half of 
the squared geodesic distance between two points with the conformal vectors
$\vec x_1$ and $\vec x_2$\ .

Thus it is necessary to obtain an approximate expression for $G^{(2)}$ 
which is valid for a small enough $\sigma$.
The well known way to obtain the desired result is the use of the 
Riemannian coordinates. Let us choose the origin of Riemannian
frame at the centre of the geodesic, connecting the points $\vec x_1$
and $\vec x_2$\ , and let $\vec x$\ be the conformal radius of this
central point and  $t^a$ be a tangent vector to the geodesic at
the point $\vec x$ directed towards the point $\vec x_2$\ . 
Proceeding as in [17], we can write
\be
x^a_2=x^a+\sqrt{\frac{\sigma}{2}}t^a-\frac{\sigma}{8}
\Biggl(\nabla^a\Omega-2t^at^b\nabla_b\Omega\Biggr)+
\frac{(2\sigma)^{3/2}}{48}\Biggl(t^a\Bigl(t^bt^c\nabla_{bc}\Omega+
\ee
$$(t^b\nabla_b\Omega)^2-\frac{1}{4}\nabla_b\Omega\nabla^b\Omega\Bigr)-
\frac{1}{2}\nabla^a(t^b\nabla_b\Omega)-\frac{1}{2}t^b\nabla^a_b\Omega\Biggr)+
O(\sigma^{5/2})\ .$$
Corresponding equation for $x^a_1$\ can be obtained from (20) 
by changing the sign before the vector $t^a$\ .

>From (20) we obtain that the Euclidean norm
of the vector $(\vec x_1-\vec x_2)$\ has the following form
\be
|\vec x_1-\vec x_2|^2=e^{\Omega(\vec x)} 
2\sigma(\vec x_1, \vec x_2)\Biggl(1+\frac{\sigma}{12}
t^at^b(\nabla_a\nabla_b\Omega+\nabla_a\Omega\nabla_b\Omega-
\ee
$$\frac{1}{2}\gamma_{ab}\nabla_c\Omega\Delta^c\Omega)\Biggr)+
O(\sigma^3)\ .$$

Substituting this expression into (17) and supposing $\sigma$ to be 
small enough we get
\be
G^{(2)}(\vec x_1, \vec x_2)=G_E^{(2)}(\vec x_1, \vec x_2)+
\frac{\Omega(\vec x)}{4\pi}+
\ee
$$\frac{\sigma}{48\pi}
t^at^b\Biggl(\nabla_a\nabla_b\Omega+\nabla_a\Omega\nabla_b\Omega-
\frac{1}{2}\gamma_{ab}\nabla_c\Omega\nabla^c\Omega\Biggr)+
O(\sigma^2)\ .$$

>From the last expression we can obtain 
for the coincidence limit of the regularized value
of the Green function the following result
\be
G^{(2)}_{reg}(\vec x, \vec x)=
\lim\limits_{\vec x_1, \vec x_2\to\vec x}
\Biggl(G^{(2)}(\vec x_1, \vec x_2)-G_E^{(2)}(\vec x_1, \vec x_2)\Biggr)=
\frac{\Omega(\vec x)}{4\pi}\ .
\ee
The result is expressed in terms of conformal factor only, which depends
on the global structure of the surface. So we can say that the 
coincidence limit of the regularized Green function plays the role of
the local test
of the global structure of the surface.

There are some ways to obtain 
the regularized value of the derivative
of $G^{(2)}(\vec x_1, \vec x_2)$\ .
One can start from
the explicit expression (17) and after the differentiation 
with respect to $x_1$ proceed
along the same lines as above,
or we can use the decomposition (22).
In this last case we must subtract $G^{(2)}_E$
from (22), and after that, perform the differentiation. If we 
intend to differentiate with respect to the coordinates of the
points $\vec x_1$\ or $\vec x_2$\ , it is necessary to remember that all
the tensor quantities in (22) are taken at the middle point and that the
coordinates of this middle point are functions of $\vec x_1$
and $\vec x_2$\ . To take into account this dependence let us
return to the Eq.(20). This equation gives us that
\be
x^a=\frac{1}{2}(x_1^a+x_2^a)+\frac{\sigma}{8}\Biggl(\nabla^a\Omega-
2t^a t^b\nabla_b\Omega\Biggr)+O(\sigma^{2})\ .
\ee
>From this equality
one can obtain 
\be
\frac{\partial x^a}{\partial x_1^b}=\frac{1}{2}-\frac{\sqrt{2\sigma}}{8}t_b
\Biggl(\nabla^a\Omega-2t^at^c\nabla_c\Omega\Biggr)+O(\sigma)\ .
\ee
Differentiating (22) with respect to $x_1$\ and taking into account
(25) we get
\be
\frac{\partial}{\partial x^a_1}\Biggl(G^{(2)}(\vec x_1, \vec x_2)-
G_E^{(2)}(\vec x_1, \vec x_2)\Biggr)=\frac{\nabla_a\Omega}{8\pi}+ 
\ee
$$\frac{\sqrt{2\sigma}}{48\pi}t^a\Biggl(\nabla_b\Omega\nabla^b\Omega-
2(t^b\nabla_b\Omega)^2+t^bt^c\nabla_{bc}^2\Omega\Biggr)+
O(\sigma)\ .$$

>From the last expression 
we obtain a very interesting correspondence 
\be
\lim\limits_{\vec x_1, \vec x_2\to \vec x}\frac{\partial}
{\partial \vec x_1}\Biggl(G^{(2)}(\vec x_1, \vec x_2)-
G^{(2)}_E(\vec x_1, \vec x_2)\Biggr)=
\ee
$$\frac{1}{2}\frac{\partial}{\partial\vec x}
\lim\limits_{\vec x_1, \vec x_2\to \vec x}
\Biggl(G^{(2)}(\vec x_1, \vec x_2)-
G^{(2)}_E(\vec x_1, \vec x_2)\Biggr)\ ,$$
which shows that the coincidence limit and the differentiation 
with respect to coordinate are noncomutative operations.

So, the regularized value of the derivative is equal to
\be
\Biggl(\frac{\partial}{\partial x^a}G^{(2)}\Biggr)_{reg}(\vec x, \vec x)=
\frac{\nabla_a\Omega(\vec x)}{8\pi}\ .
\ee

It is clear that the same procedure permits us to calculate the
coincidence limit of all the other derivatives of the Green
function. It may be necessary if we want to consider 
the self-forces on a current distributions with some inner
structure. 

\section{\bf Explicit solution for topological self-force 
on multistring spacetime}

The results obtained in the previous sections enable us to conclude
the considerations on the problem of topological self-action on a
linear distribution of four-current placed in the
multistring spacetime and even to consider a more general case of
the spacetime with a metric of the form (1).

Indeed from (9) and (23) we can obtain the expression for the 
regularized electromagnetic self-energy per unit length of the current
\be
\frac{E_{stat}^{reg}}{\int dz} = 
-\Omega (\vec x)\frac{\left( J_t^2+J_z^2 \right)}{2}\ .
\ee
The obtained
value depends on the coordinates on the section $t=const, z=const$\ . This is a 
trivial consequence of the
absence of translational invariance in the direction perpendicular to 
the $z$-axis and means that the current must feel the action of
some self-force.
As it was stressed in Sec.2
to obtain the expression for the regularized self-force we must start
from (13) and after the use of (28) we obtain that the nonvanishing components
of the force are as follows  
\be
\frac{F^a (\vec x)}{\int dz} = 
\frac{J^2}{2}\gamma ^{ab} \partial _b \Omega(\vec x)\ .
\ee
>From the last expressions for the the observer in the locally Cartesian
frame we get
\be
\frac{\vec F}{\int dz } = \frac{J^2}{2}  e^{\frac{\Omega}{2}}
\vec \bigtriangledown
\Omega \ .
\ee

Let us apply the above results and consider a particular 
and a very interesting case 
of the multiconical space, when
the subspace $t=const, z=const$\ is a locally flat two-dimensional surface with
a number of conical singularities located at the points with conformal
radii $\vec x_i$\ . For these purposes the conformal factor
in (14) must be taken in the form [20, 21]
\be
\Omega(\vec x) =  \sum _{i=1}^{N} 2(1 - b_i )  \log |\vec x -\vec x_{i} | \ .
\ee
In the expression above $b_i$\ stands for the parameter which determines
the angle deficit of the $i$-th conical singularity.
>From the cosmological point of view this solution
describes the ultrastatic spacetime of $N$\ parallel cosmic
strings [20]. In the same context, the current 
(2) may be associated with the current of superconducting cosmic
strings which were predicted by Witten [22].

>From (29) and (32) we see that
the contributions to the 
self-energy from different singularities add to each other
\be
\frac{E_{stat}}{\int dz}=-(J_t^2+J_z^2)\sum\limits_{i=1}^{N} 
(1-b_i)\log |\vec x-\vec x_i|\ .
\ee

One can conclude that there is a principle of superposition and
that the addition of an extra string does not lead to an extra
difficulty.

It is easy to understand that the real situation is much more
complicated.
One has to remember that $\vec x_i$\ in (32) are not the real geodesic 
distances from the corresponding singularities but the conformal
radii. So, we obtained a superposition principle but in terms of
the Euclidean plane, with which our two-dimensional multiconical
subspace is conformally connected. When the self-energy is
expressed in terms of geodesic distances from the strings it is a
much more complicated function, and there is no superposition
principle in the explicit sence of this word. 

To illustrate this statment let us consider two-strings spacetime and
a current between them (both strings and the current 
lie in one and the same plane 
$x_2=0$)\ . Suppose for simplicity that the strings have equal
tensions.
Direct calculation gives that in this case the difference between the explicit
expression for the topological self-energy 
and the one obtained under the assumption that
the superposition principle holds reads
\be
\frac{E_{stat}-E_{sup}}{\int dz}=-(1-b)(J_t^2+J_z^2)F(x)\ ,
\ee
where
\be
F(x)=\log\frac{x(1-x)}{(\rho_1\rho_2)^{\frac{1}{b}}}\ ,\qquad 0<x<1\ .
\ee
The coordinate $x$ indicates the position of a current
between the strings, and $\rho_{1, 2}$
are the dimensionless geodesic distances from corresponding string to the current
\be
\rho_1=\int\limits_0^x \frac{dx}{x^{1-b} (1-x)^{1-b}}\ ,\qquad
\rho_2=\int\limits_x^1 \frac{dx}{x^{1-b} (1-x)^{1-b}}\ .
\ee

Analytic estimates show that the value
of $F(x)$ decreases with $b\rightarrow 1$
as $(1-b)$. So for the cosmic string network $((1-b)\sim 10^{-5})$
we indeed can use the
superposition principle, but it is not the case for the disclinated
medium where $(1-b)\sim 1$\ .

Let us consider the self-force now.
>From (31) it is very easy to show that in the locally inertial frame 
the force of topological self-action has the form
\be
\frac{\vec F (x) }{\int dz } = e^{\frac{\Omega }{2} }
 \sum _{i=1} ^{N}   J^2 (1-b_i)
\frac { \vec n}{| \vec x -\vec x_i | }  , \ \ \mbox{where} \ \
\vec n = \frac {\vec x - \vec x_i}{| \vec x - \vec x_i |}\ .
\ee
We see again that because of the factor
$e^{\Omega/2}$\ the contributions from different strings are
mixed and the force which is measured by a locally inertial observer
is not a sum of the contributions from different strings.

To interprete the obtained result, let us consider the simplest case of
only one conical string ($N=1$)\ stretched along the $z$ axis. 
Introducing a new radial
coordinate $\rho = b^{-1}|\vec x|^{b} $\ , we get
\be
\frac{\vec F (x) }{\int dz } =
J^2 \frac{(1-b)}{b\rho ^2 } \vec \rho\ .
\ee
This new coordinate $\rho $ gives the true geodesic distance from the
string.  Thus for the locally inertial observer the force on the
four-current is the same as the force between two parallel four-currents, our
one and
$I^{\mu}(x)=(2b)^{-1} (1-b) J^{\mu }$\ ,
in Minkowski space.
The sign of the force depends on the sign of $J^2$\ . For $b<1$ the
force is attractive for the space-like current, and repulsive for
the time-like one.

We see that if
$N=1$ our result coincides with the result [15], but it should be pointed out
that the method of separation of variables used in [15], can not be
applied in the multistring case, and of course, it can not be used in
more general cases.

To conclude this section let us add some discussion concerning the
connection of the obtained results and the symmetries of 
the system under consideration. 

The expression for the self-force in the multistring space (37) is
proportional to the squared invariant amplitude of the current
$J^2$\ , and so, the force is independent of an observer's speed
along the strings. We shall show that this 
fact can be explained on the pure qualitative level, and
moreover for the case of only one conical string the
expression for the self-force can be obtained up to a numerical coefficient.

Let us remember that the vector $z\partial/\partial t + t\partial/\partial z$\ 
is the
fourth of the four Killing vectors of the locally flat conical space
mentioned in the Introduction. Indeed,  
in the case of a straight string such a Lorentz transform 
changes the parametrization of its world sheet, but the 
Nambu action is invariant under general reparametrizations of the
world sheet [3, 4].
The existence of this symmetry means that the
longitudinal motion of a Nambu string is unobservable.  

To obtain a time independent infinitely thin current we can use any
action for the superconducting string, say the Nielsen-Olesen one [23]. 
But the action for the superconducting string is invariant under
the reparametrization of the string's world sheet too.
Thus, if we consider two parallel strings, and one of the strings is a
superconducting string, a Lorentz boost along them leads to the
reparametrization of the both world sheets, and so, must be 
unobservable. This means that the force may depend on the "invariant
charge" $J${\footnote{\normalsize Similar result was
obtained under consideration of radiation prosesses in the
cosmic string network [19]}}\,
and the only possibility to obtain the expression
for the force with the correct dimension is to write
$F\sim J^2/\rho$\ but this is the force between two parallel
four-currents which are proportional to each other. 

Then, we consider the interaction of a conical gravitational
field with the electromagnetic field only. So, we can extend 
the qualitative result above to any current independently of
the nature of its carrier.

We see that the result (38) can be predicted on a pure qualitative 
level. 
But it is not the case in the spacetime of multiple cosmic
strings because in this case we have several parameters (conformal
distances between the strings) with the same dimension.
This makes it impossible to obtain the result (37) by a qualitative 
discussion 
along the same line as above.

\section{\bf Conclusion}

Spaces with conical singularities 
have
a lot of physical applications in the phenomena of very different
spacetime scales. In some cases the multistring solution
corresponds to the real physical situation much more
than the spacetime of only
one cosmic string. 

Our primary goal was to show that the naive point of view that
some results concerning the behaviour of classical or quantized 
matter on 
the multistring space can be obtained as a 
superposition of individual contributions from the separate strings 
is wrong. Multiconical boundary conditions change the result
essentially, and one can speak about the 
superposition principle
approximately in a very limited number of cases.

In the paper we have considered an explicitly solvable problem
which enables us to establish certain nonlocal effects in the 
spacetime of multiple cosmic strings. To our mind this is the
first explicitly solvable problem in the 4D locally flat
multiconical space that have been considered in 
scientific literature.

We hope that the results presented here can be of interest 
in investigating the behaviour of superconducting strings in
the cosmic string network and the topological effects in
disclinated media. 
>From the observational point of view 
the effect of topological self-action
may be of much more interest for the solid state physics because it
is proportional to the parameter $ (1-b) $, which is
very small for the GUT strings,  but may be of the order
of unity in the case of disclinations.
Our results show that in this last
case the collective nonlocal contribution to the effect which is due to 
the presence of more than one conical singularity may be of
great importance and must be taken into account.\\

{\bf Acknowledgments}

Two of us (E.R.B.M) and (V.B.B) would like to thank CNPq 
for the partial financial
support.

One of the authors (Yu.V.G.) is grateful to Department of Physics of
the Federal University of Paraiba (Brazil) where this work was
performed, for kind hospitality.
His work was supported in part by CAPES and in part by 
the Russian Foundation for Basic
Reseach, grant 96-02-18899 and grant
97-02-18003.

\end{document}